\documentstyle[prl,aps,multicol,epsf]{revtex}
\begin{document}
\draft
\widetext

\title{Distribution function of mesoscopic hopping conductance}

\author{Liqun He${}^{1,2}$, Eugene Kogan${}^{1,3}$, Moshe Kaveh${}^{1,3}$, Shlomo 
Havlin${}^1$, and Nehemia Schwartz${}^1$,}

\address{${}^1$Minerva Center and Jack and Pearl Resnick Institute of Advanced 
Technology,\\
Department of Physics, Bar-Ilan University, Ramat-Gan 52900, Israel}

\address{${}^2$Department of Thermal Science and Energy Engineering,\\ 
University of Science and Technology of China, Hefei, P.R.China}

\address{${}^3$ Cavendish Laboratory, Madingley Road, Cambridge CB3 OHE, UK}
 
\date{\today }

\maketitle
\widetext
\begin{abstract}
\leftskip 54.8pt
\rightskip 54.8pt

We study by computer simulation distribution functions (DF) of mesoscopic 
hopping conductance.  The DFs obtained for one-dimensional 
systems were found to be quite close to the predictions of the theory by Raikh 
and Ruzin. 
For $D=2$, the DFs both for  narrow system and thin film look similar (and close 
to the $1D$ case).The distribution function for the conductance of the square 
sample is nearly Gaussian.   

\end{abstract}


\begin{multicols}{2}
\narrowtext

Mesoscopic conductance fluctuations in the insulating regime  of
small, disordered transistors were first observed by Pepper 
\cite{pep} in GaAs MESFETs and then studied in detail in Si
MOSFETs  by Fowler, Webb and coworkers \cite{fow} in the early
1980s. Extremely  strong random fluctuations, spanning several
orders of magnitude,  were observed at low temperatures in the
conductances of narrow-channel  devices as the gate voltage was
varied. The   explanation
was provided by Lee \cite{lee} who proposed a model in which 
electrons move by variable-range hopping (VRH) along a
one-dimensional  (1D) chain. A number of elementary hopping
resistances, each  depending exponentially on the separation and
energy difference  between sites, are added in series to give
the overall resistance  of the chain. In this model it is
assumed that, because of the  extremely broad distribution of
the elementary resistors, the  total chain resistance can be
well approximated by that of the  single most resistive hop. The
fluctuations then arise as a consequence  of switching between
the pairs of localized sites responsible  for the critical hop
as each elementary resistance reacts differently  to a change in
the chemical potential. These fluctuations are  therefore of
``geometrical'' origin, arising from the random  positioning of
localized sites in energy and space, as distinct  from the
``quantum'' nature of the tunneling mechanism which  would be
strongly affected for example by an applied magnetic  field.
Serota, Kalia and Lee \cite{ser} went on to simulate the
ensemble  distribution of the total chain resistance  $R$ and
its dependence on the temperature  $T$ and the sample length
$L$. In their ensemble, the random impurities are distributed 
uniformly in energy and position along the chain. In
experiments  a single device is generally used, so that the
impurity configuration  is fixed, and fluctuations are observed
as a function of some  variable external parameter such as the
chemical potential. An  ergodicity hypothesis is then invoked to
the effect that the  same ensemble is sampled in both cases,
something that has been  verified experimentally by Orlov  {\it
et al.} \cite{orl1}. Using the natural logarithm of the
resistance,  the authors of Ref. \onlinecite{ser} 
obtained for the mean and standard deviation:

\begin{equation}
\label{1}
\langle \ln\rho \rangle \sim \left(\frac{T_0}{T}\right )^{1/2}
\left [\ln\left(\frac{2L}{\xi}\right)\right]^{1/2}
\end{equation}
\begin{equation}
\label{2}
s\equiv\langle \left(\ln\rho- \langle \ln\rho \rangle\right )^{2}
\rangle \sim \left(\frac{T_0}{T}\right )^{1/2}
\left [\ln\left(\frac{2L}{\xi}\right)\right]^{-1/2}
\end{equation}
where $\xi$ is the localization radius and $T_{0}$ is the
characteristic temperature for Mott VRH:  $T_{0}~=~1/k_{B}\rho
\xi$ ($\rho$ is the density of states at the Fermi energy). It
can  be seen that the size $s$ of the fluctuations decreases
extremely slowly with  length, a result characteristic of 1D
which was first pointed  out by Kurkijarvi \cite{kur}. The
explanation is simply that exceptionally  large resistance
elements, even though they may be statistically  rare, dominate
the overall resistance since they cannot be by-passed  in this
geometry. The averaging assumed in the derivation of  Mott's
hopping law for 1D does not occur and the total resistance 
takes on the activated form of the largest individual element.

A detailed analytical treatment of this model was undertaken  by
Raikh and Ruzin (RR) \cite{rai1,rai2} who divided the problem  into
a number  of length regimes. Their theory introduces the concept
of the  ``optimal break'', the type of gap between localized
states (on  an energy versus position plot) which is most likely
to determine  the overall resistance. The optimal shape of such
a state-free  region has maximal resistance for the smallest
area and turns  out to be a rhombus. A sufficiently long chain
will have many  such breaks in series to give a most probable
resistance

\begin{equation}
\label{eqnR2}
R=R_0\frac{L}{\xi}
\left(\frac{T_0}{T}\right )^{1/2}\exp
\left(\frac{T_0}{2T}\right )^{1/2}
\end{equation}
where $R_{0}$ is the prefactor in the Mott VRH formula, and $\rho=R/R_0$. This
formula  is valid  only for substantially long samples. More exactly, it is 
valid when $\nu \gg 1$, where $\nu$ is 
a parameter defined implicitly by

\begin{equation}
\label{eqnNu}
\nu= \frac{2T}{T_0}\ln \left(\frac{L\nu^{1/2}}{\xi}\right).
\end{equation}

When the expected number of optimal breaks in the chain 
becomes  of order one ($\nu \leq 1$), which corresponds to the normal 
experimental 
situation, the resistance of the chain is determined by a few 
sub-optimal breaks of which the expected number occurring in  a
chain of length $L$ is approximately one. In this case, the most probable value 
of the
resistance  (or its logarithm  $Q=\ln\frac{R}{R_0}=\ln\rho$) is
\begin{equation}
\label{eqnR1}
Q=\frac{\nu^{1/2}T_0}{T}\approx
\left\{2\frac{T_0}{T}\ln\left[ \frac{L}{\xi}
\left(\frac{T_0}{T}\right )^{1/2}
\ln^{1/2}\left(\frac{L}{\xi}\right)\right]\right\}
\end{equation}

The probability distribution function (DF) for the quantity 
$f(Q)$ is best written in terms of $\nu$ and a new parameter
$\Delta$
\begin{equation}
\label{7}
\Delta= Q-\frac{\nu^{1/2}T_0}{T},
\end{equation}
For $\nu <1$ it is given 
by the following integral:
\begin{eqnarray}
\label{eqndf}
f(\Delta)=\frac{e^{\Delta}}{\pi}\int^{\infty}_0
\exp\left(-x^{\nu^{1/2}}\cos\frac{\pi\nu^{1/2}}{2}\right) \nonumber\\
\cdot\cos\left(xe^{\Delta}-x^{\nu^{1/2}}\sin\frac{\pi\nu^{1/2}}{2}\right)dx, 
\end{eqnarray}
where
$f(\Delta)$ is a function with a peak close to $\Delta = 0$
 and width 
determined by $\nu$. 
There is a simple relationship 
between $\nu$ and the variance of $Q$:
\begin{equation}
\label{eqnvar}
\langle Q^2 \rangle - \langle Q \rangle^2=\frac{\pi^2}{6}
\left(\frac{1}{\nu}-1\right)
\end{equation}
This theory is equally applicable \cite{rai3} to the case of the
transverse  conductance $\sigma$ of a thin film or barrier. Instead
of a sum of series  resistances, the required quantity is the
sum of parallel conductances  representing conducting chains of
hops traversing the film. Whereas  in 1D the total resistance is
determined by the blocking effect  of the critical hop, here the
total conductance is dominated  by an optimal ``puncture'': an
uncommonly high-conductance hopping  chain through the barrier
which effectively shorts out all other  current paths. On a
logarithmic scale, since $\ln\rho=~-\ln\sigma$, the DFs for the two
geometries are simply reflections  of each other. The variation
of the width and peak position with  $\xi$ and $\rho$ is
different, however, in the two cases. 
In 1D the importance of blocking resistors adds weight  to the
contribution of extremely high resistances and produces  a long
tail out to low values of $\ln\sigma$. For a short 2D barrier the DF
has the opposite asymmetry  with a tail out to high
conductances, reflecting the effect of  punctures in shorting
out less conductive paths. In fact the  form of the DF is
universal, the theory requiring only that the  elementary
quantities to be summed are independent and come from  an
exponentially wide distribution. The microscopic details of  the
conduction mechanism enter only into the dependence of $\nu$ and
$\Delta$ on external parameters such as the temperature  and magnetic
field. The requirement of independence in the case  of the
barrier means that conductive chains must be sufficiently  far
apart, which should be satisfied for a barrier with a
sufficiently  large aspect ratio $W/L$. 
We use the description ``short 2D'' for this short-, 
wide-channel geometry to distinguish it from the square 2D geometry 
in which conduction is via an interconnected percolation network.
 
The aim of the present work is to do numerical simulation on the 1D and 2D 
mesoscopic systems in Mott hopping regime, and find the distribution functions 
of conductance in these systems.  We start by replacing the transport problem 
with a random-resistor network in which the hopping between sites $i$ and $j$ is 
equivalent to having a resistor $\rho_{ij}$ such that
\begin{equation}
\label{eqnRij}
\ln\rho_{ij}=2\alpha d +(|E_i-\mu|+|E_j-\mu|+|E_i-E_j|)/2kT ,
\end{equation}
Here $\rho_{ij}$ is the resistor between sites i and j, $\alpha$ is the inverse 
localization length,  $d$ is the distance of two localized sites, $E_i$ and 
$E_j$ are energies of site $i,j$, and $\mu$ is the chemical potential, $T$ is 
the temperature.  
Thus the mesoscopic system is reduced to a random resistor network. To find the 
resistance of the network, the resistors joining electrodes are 
selected in ascending order until the first percolation path connects the 
reservoirs.
The resistance of the percolation path is taken to be the resistance of the 
entire system.

In the simulation, first, a number of   different impurity 
configuration is generated.  Sites are randomly distributed 
along dimensions of the system 
and  their energies are chosen randomly from a uniform distribution
between $-0.5 \sim +0.5$. 
For each configuration we randomly chose the position of the chemical potential $\mu$.
Thus we can consider the chemical potential distributions (for a fixed impurity 
configuration) and the ensemble and chemical potential distributions. Typical 
results for a given configuration and given value of chemical potential are 
presented below.
For a 1D system of $L=1000$ When chemical potential $\mu=0$ and temperature 
$T=0.001$, by the method 
described above, the threshold  is  $\ln\rho_c=11.9$. Among about 500,000 
resistors, there are 53 resistors satisfying unequality 
$\ln\rho_{ij}\leq\ln\rho_c$, among which there are 12 resistors in the 
percolation path. The 
profile of their resistances is shown on Fig. 1a. We see that the 
 single largest hop in fact determines the conductance of the system. 
 With temperature increasing, the resistors in 
percolation path become closer in resistance, which can been seen on Fig. 
1b, which presents the results of similar calculations, but with  
$T=0.01$. 
\begin{figure}
\label{figratio}
\epsfxsize=1.6truein
\centerline {\epsffile {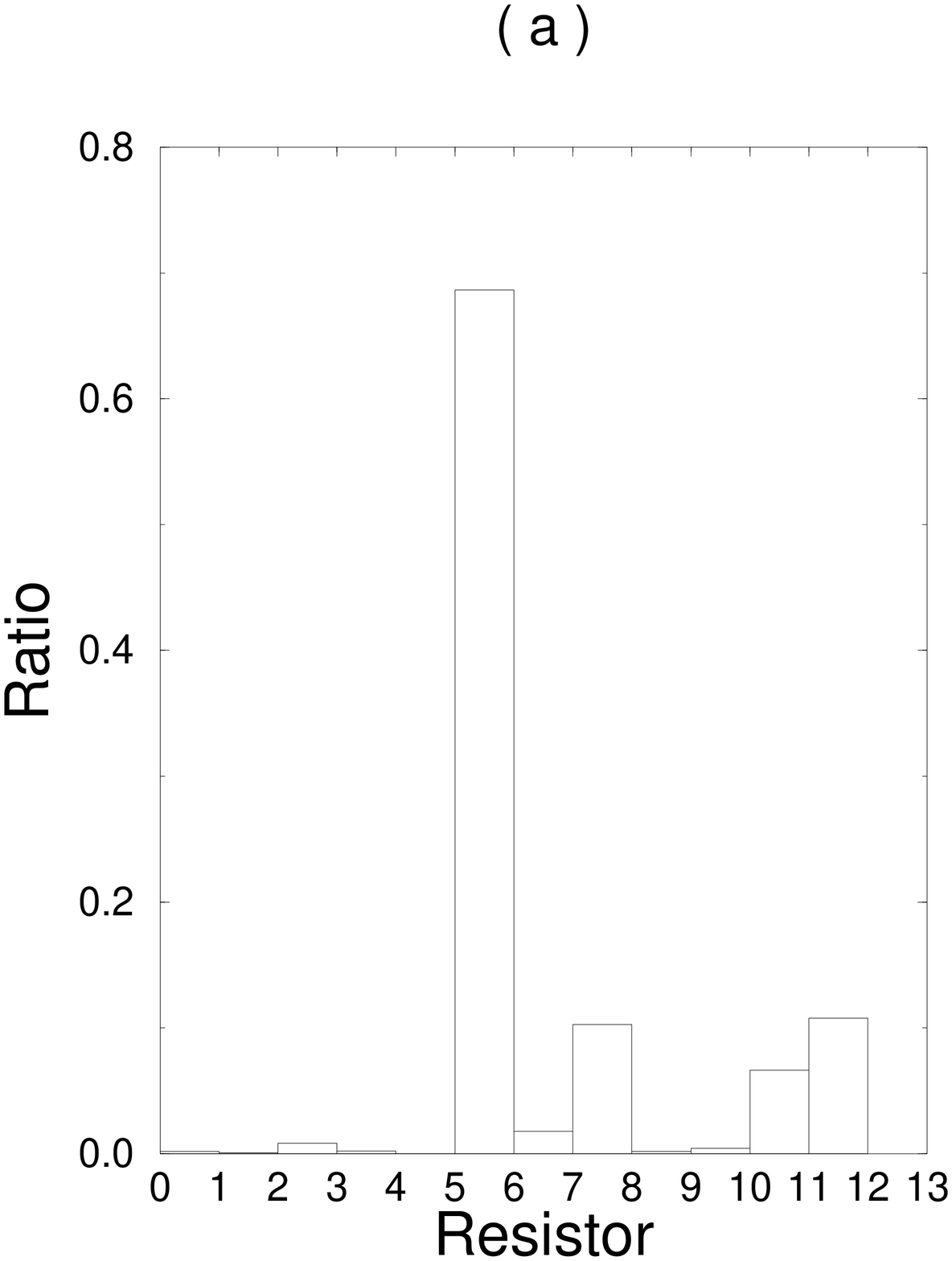}\epsfxsize=1.6truein{\epsffile{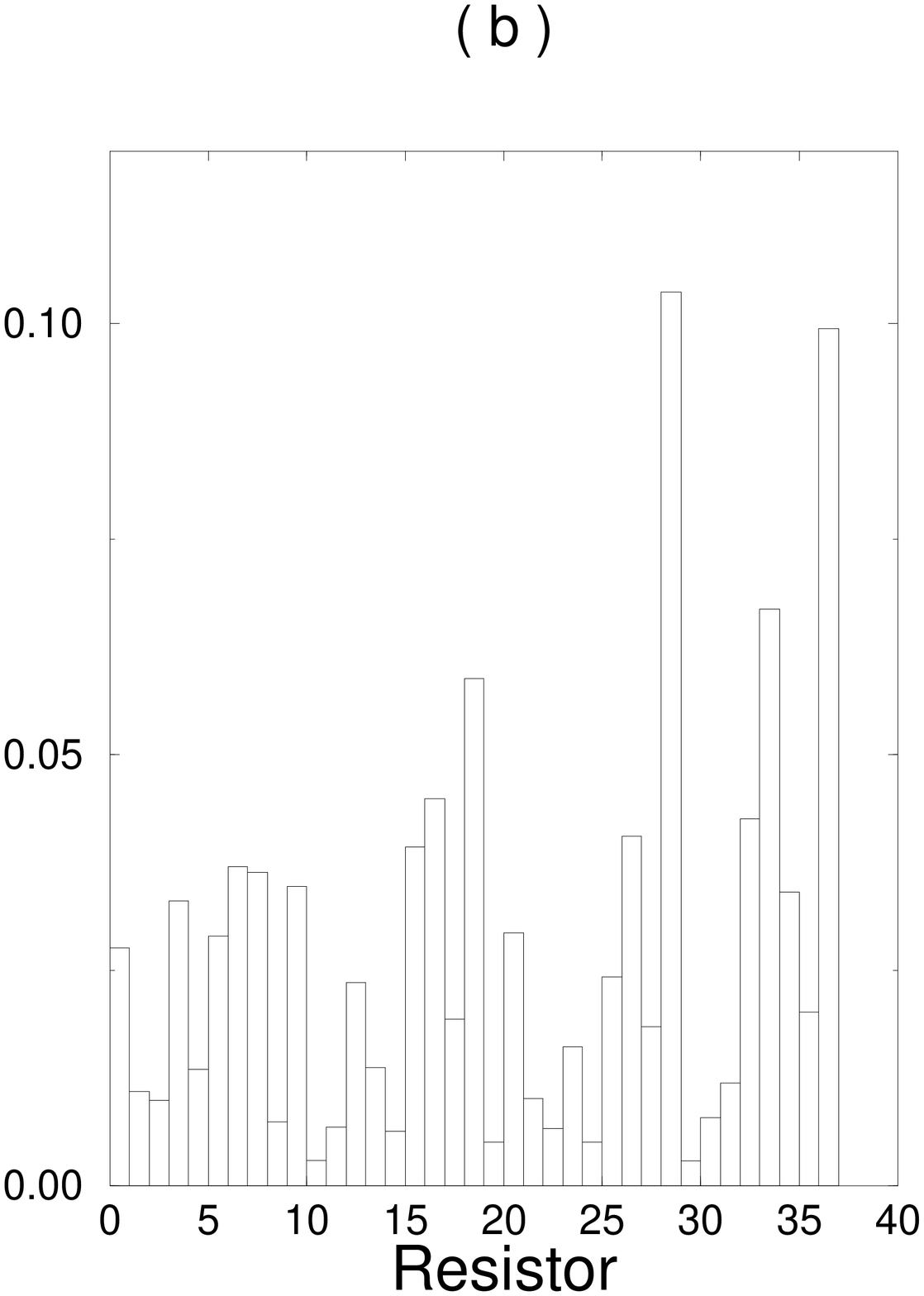}}}
\caption{The ratio of individual resistances in the percolation path to the 
path resistance: a)$T=0.001$; b)$T=0.01$.}
\end{figure}

We simulate the 1D system at first, and compare the numerical results with 
the RR theory.  The 1D system  has 
a length of 1000, and a localization length of 50. We consider three cases:

\noindent
1. $T= 0.001$, which gives  $\nu=0.225$;

\noindent
2. $T= 0.01$, which gives  $\nu=3.6$;

\noindent
3. $T= 0.015$, which gives  $\nu=5.8$.

\noindent 
For each temperature we consider 1000 ensembles, and  chemical potential range 
is $\mu=-0.1\sim+0.1$. The results of the first case is shown below.

\begin{figure}
\label{figdf}
\epsfxsize=1.6truein
\centerline {\epsffile {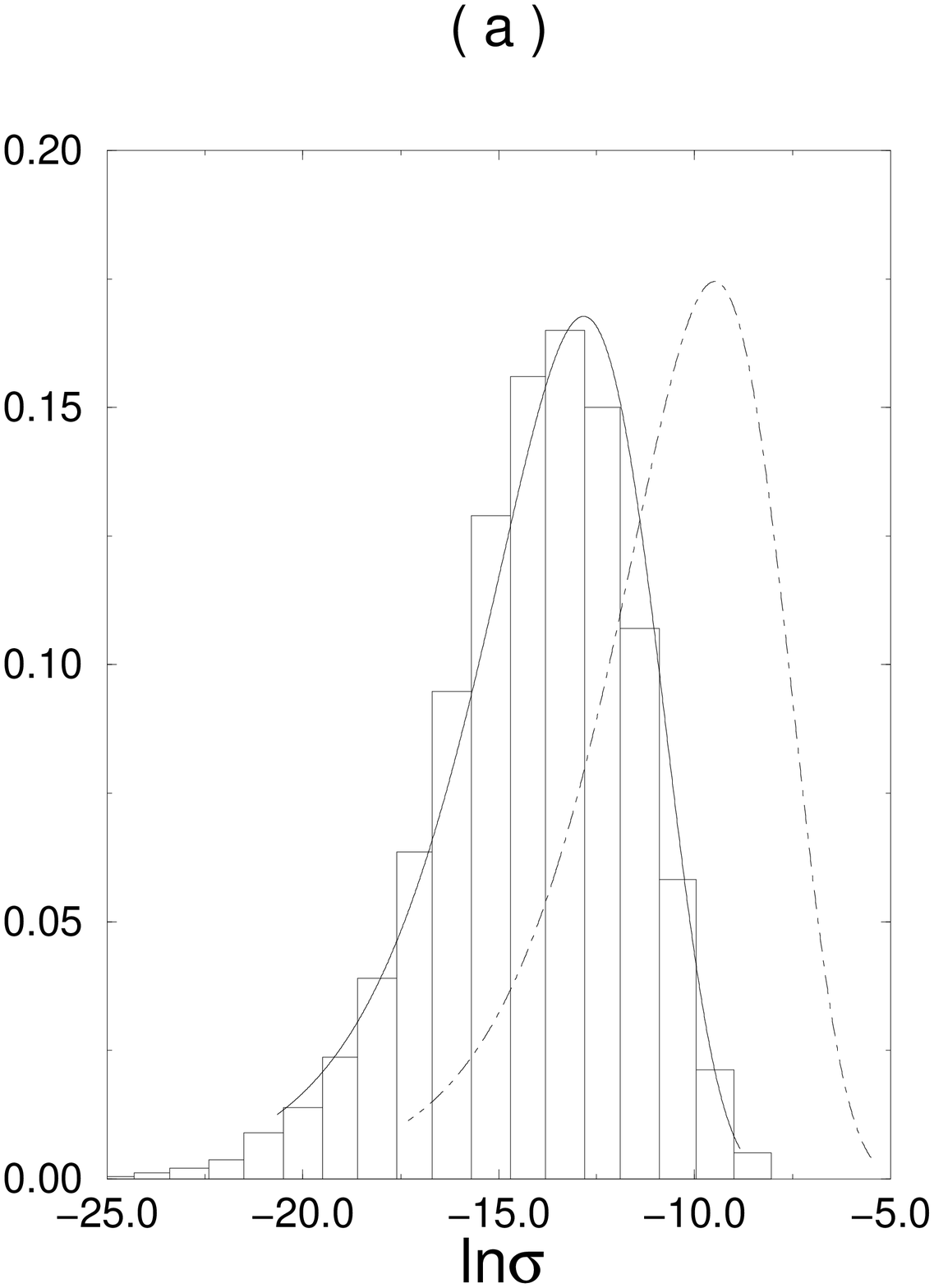} \epsfxsize=1.6truein {\epsffile{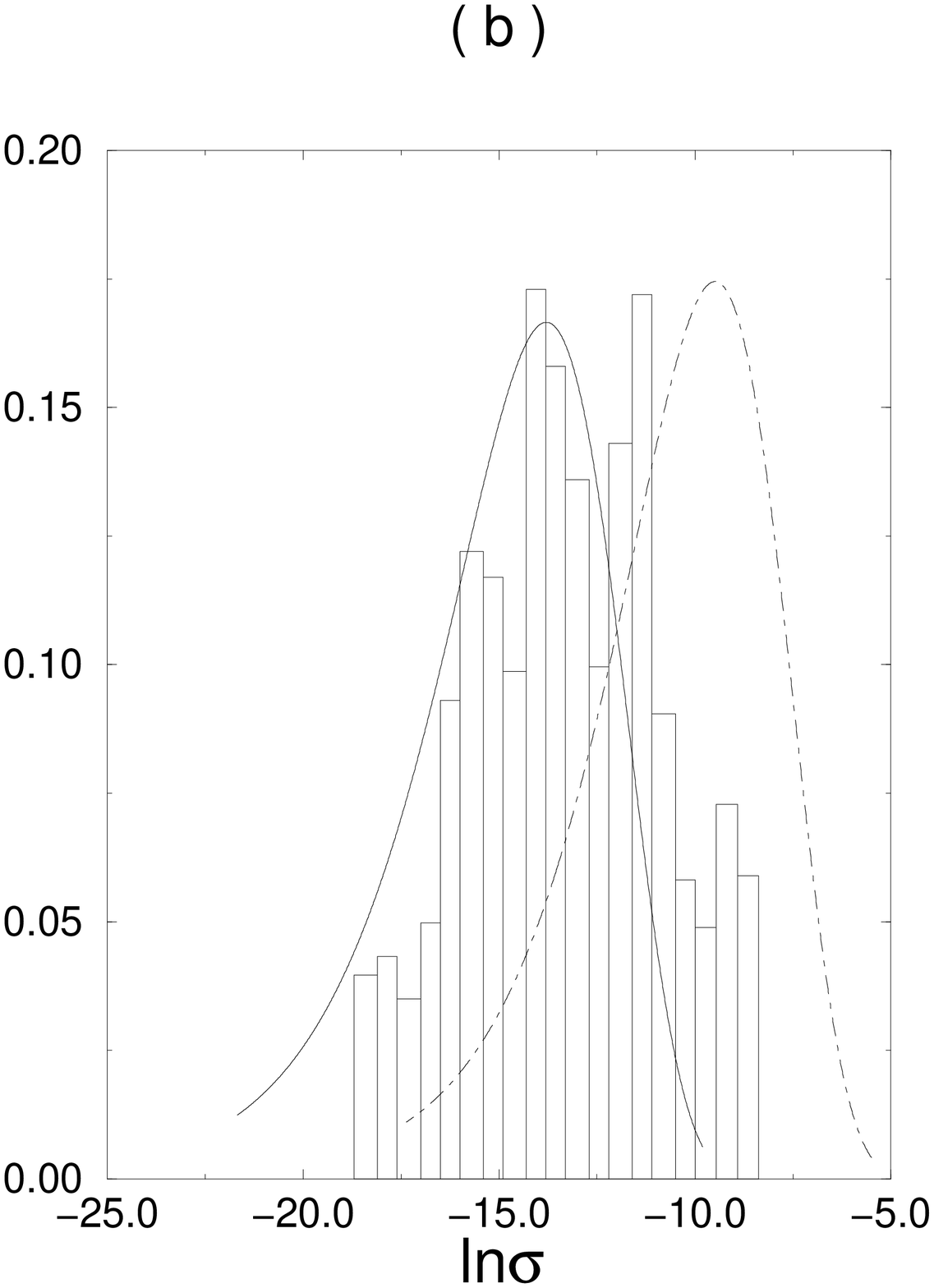}}}
\caption {The conductance  of 1D system for $\nu=0.225$: a) ensemble  
distribution function; b) chemical potential  distribution function. The 
histograms in both figures are 
the numerical result, the solid lines correspond to the fitting curves, and the 
dash-dot lines are the prediction by the RR theory.}
\end{figure}

The numerical simulation results for  $\nu=3.6$, where according to the theory,
the distribution function should be Gaussian, are presented below.
\begin{figure}
\epsfxsize = 1.6truein
\centerline {\epsffile{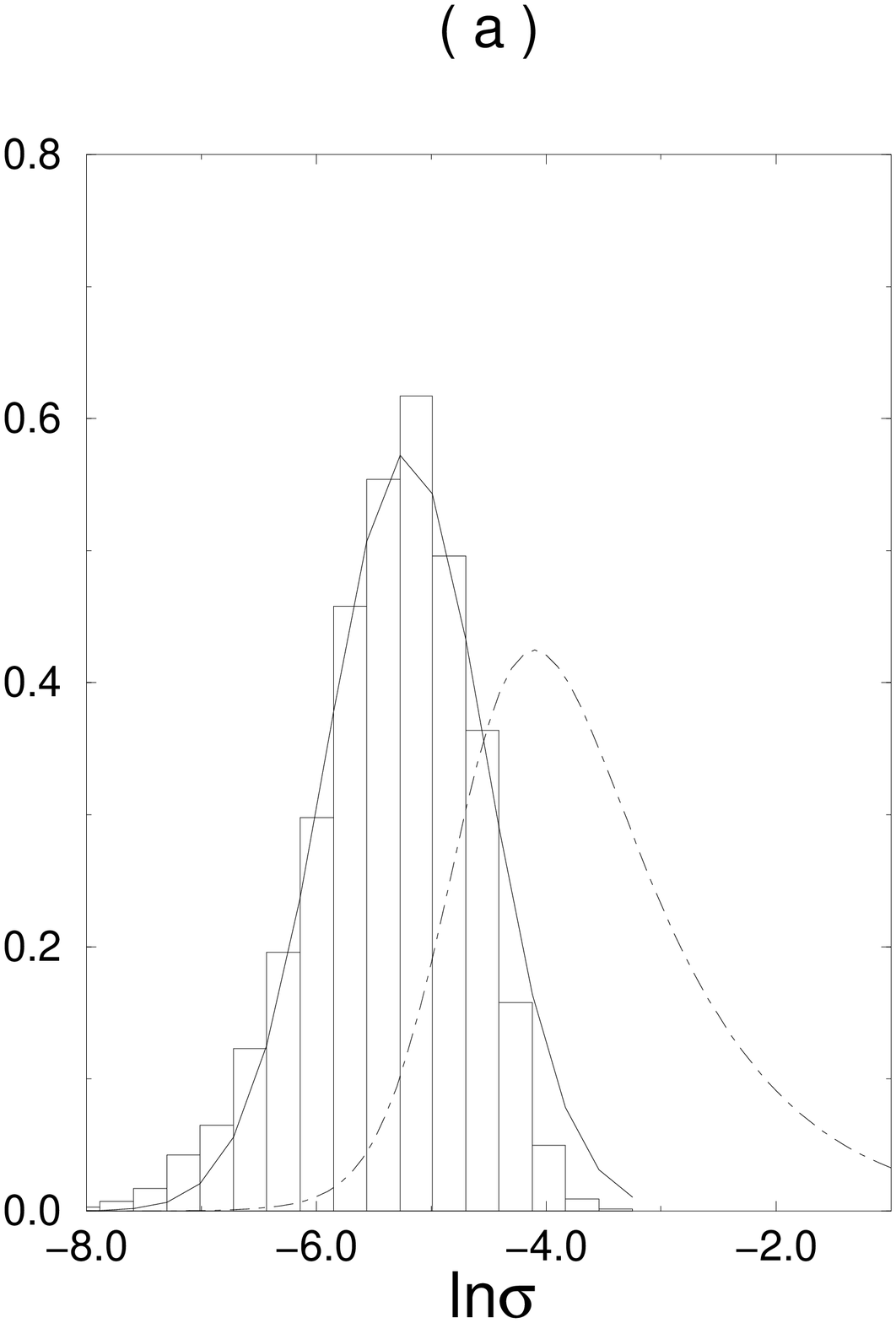}\epsfxsize=1.6truein{\epsffile{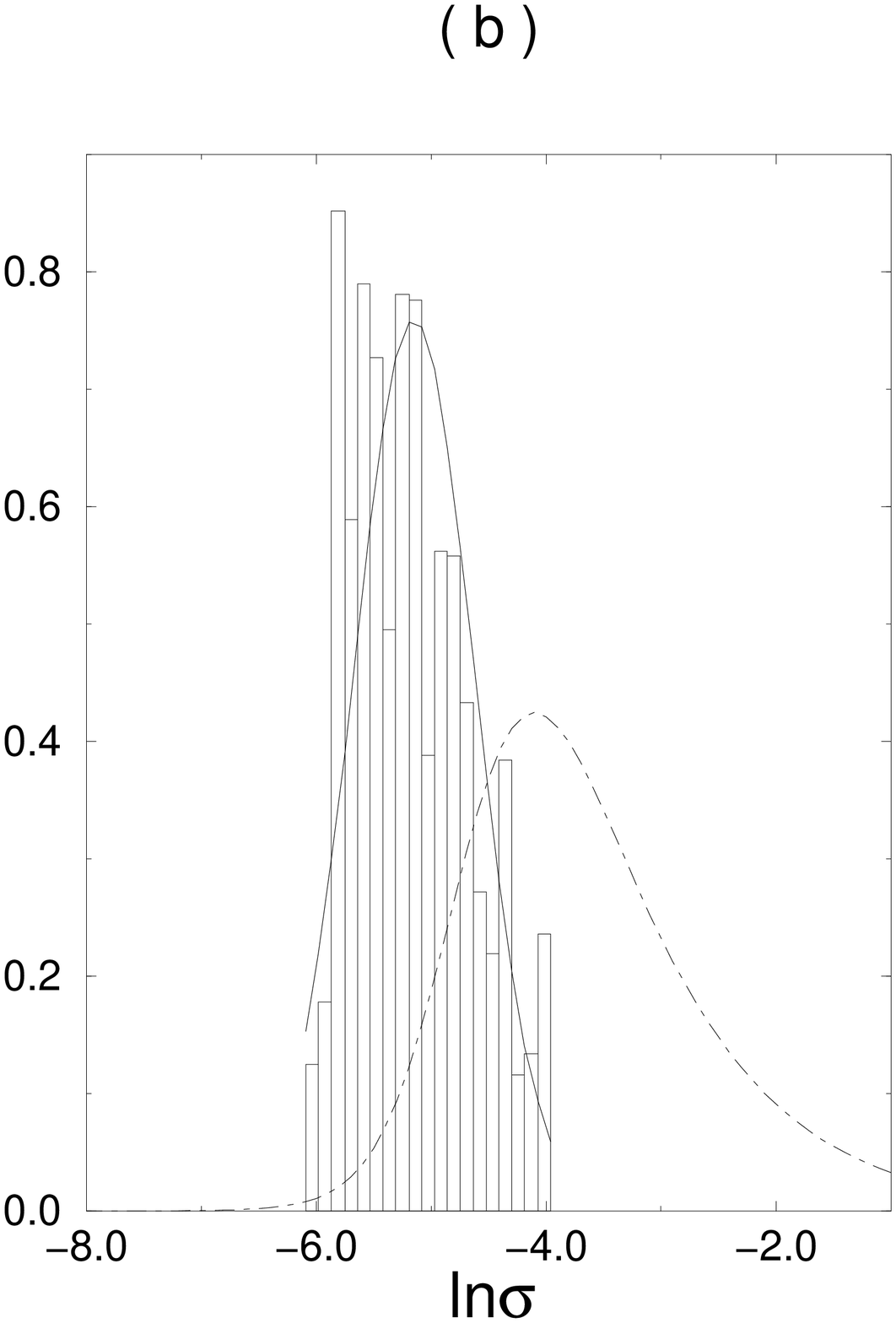}}}
\caption{The conductance  of 1D system for $\nu=3.6$: a) ensemble  distribution 
function; b) chemical potential  distribution function. The histograms in both 
figures are 
the numerical result, the solid lines correspond to the fitting curves, and the 
dash-dot lines are the prediction by the RR theory.}
\end{figure}
The result of $T=0.015$ leads to the similar figures above except for thiner 
and higher DFs, which shifts a little towards the higher conductance. 

For  2D system, we consider three particular cases:  narrow 2D system, square 
sample and thin film. 
We expect that
the  DF  of narrow 2D system should be close to that of 1D system 
and according to the RR theory, the thin film DF on a logarithmic 
scale should be a reflection of the 1D case distribution function. It is natural 
to expect that the DF of the square system would be close to a Gaussian.
For a  2D system  all 
parameters are chosen to be the same as in the 1D case, except $w=100$ in the 
case of narrow sample;
$w=1000$ and $L=100$ for thin film  and  $w=L=1000$ for square 2D 
system. The number of impurity configurations = 50 in each case.  The results 
are shown below.
\begin{figure}
\epsfxsize=1.6truein
\centerline {\epsffile{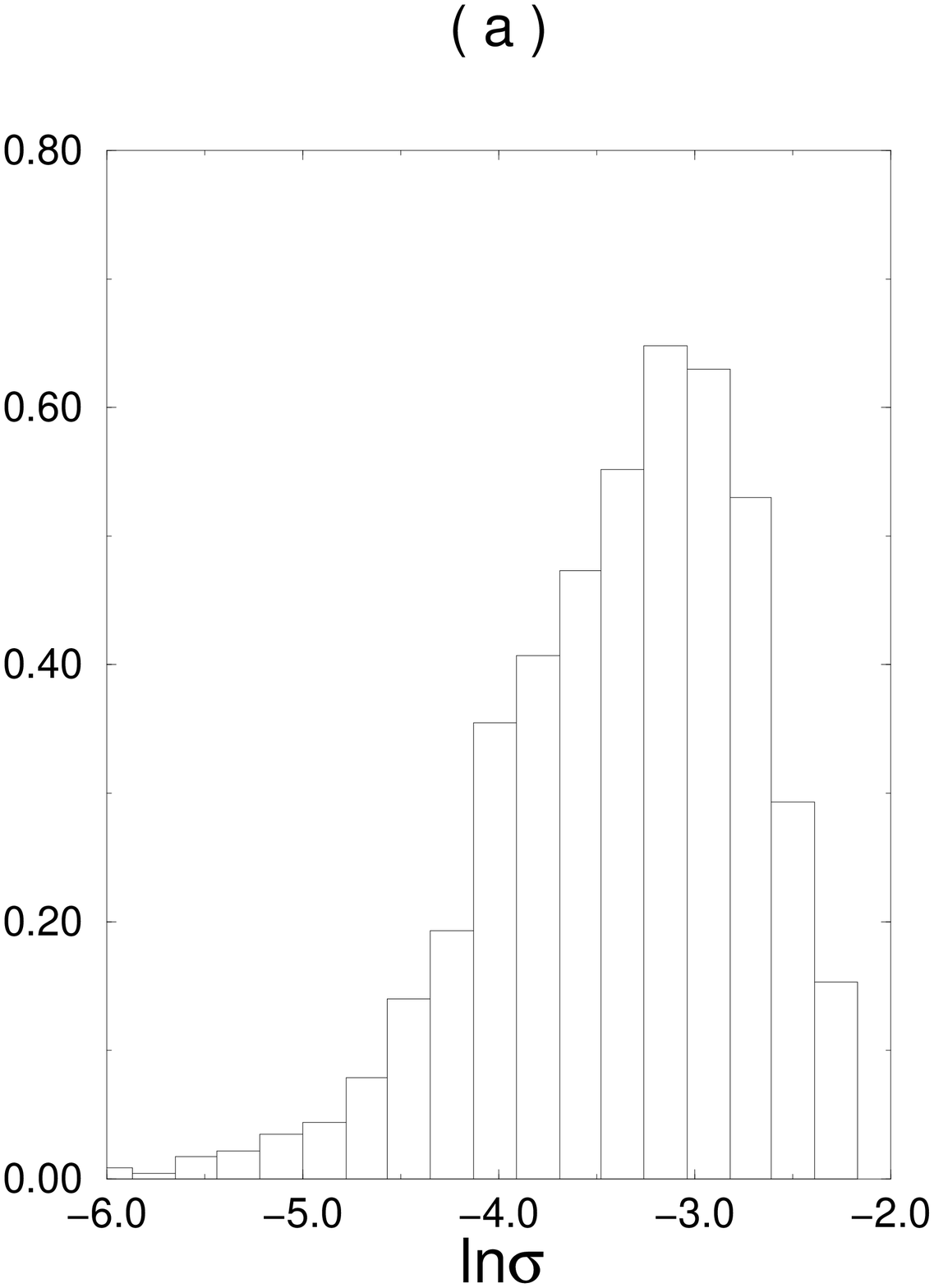} \epsfxsize=1.6truein {\epsffile{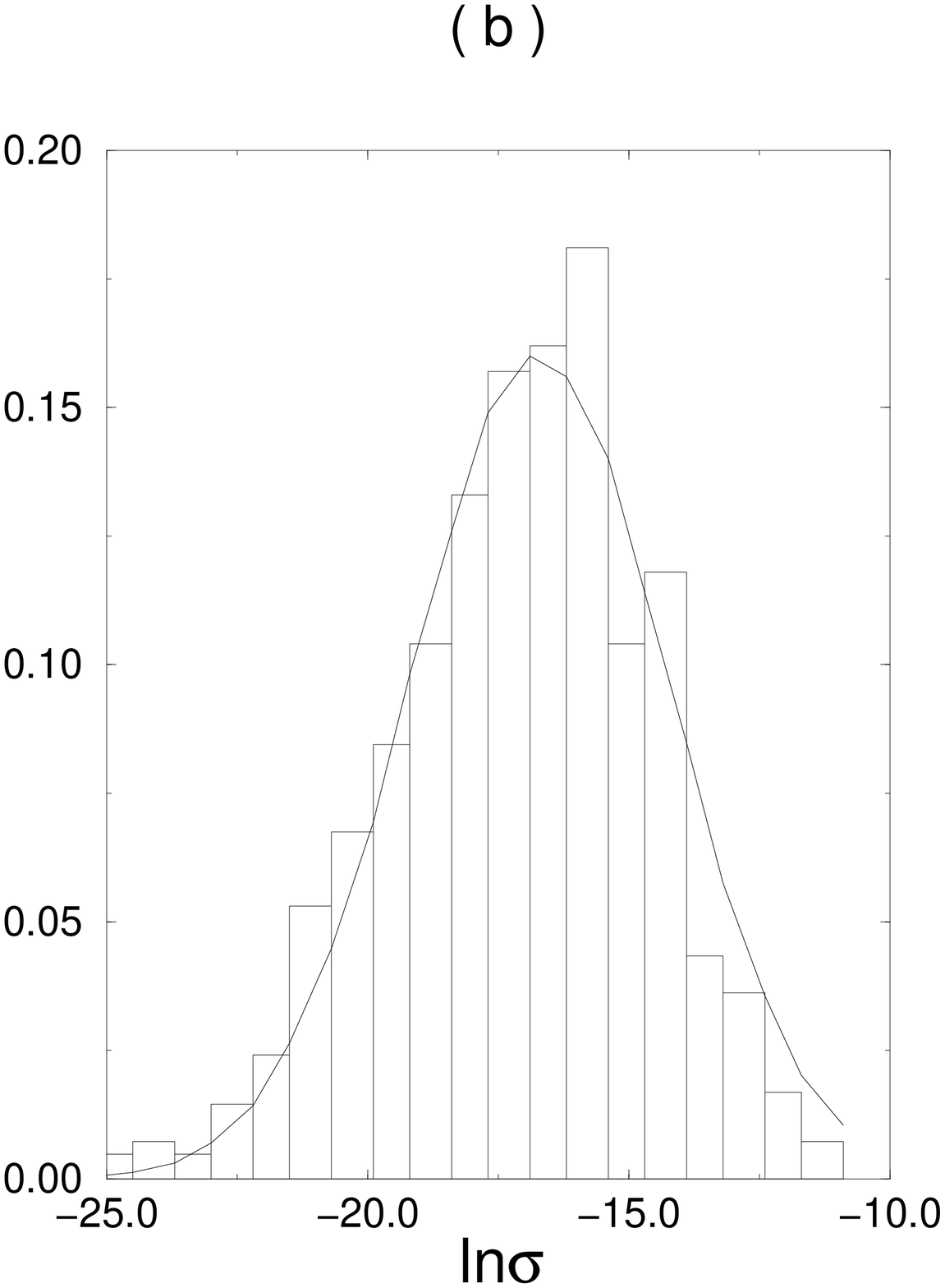}}}
\centerline {\epsfxsize=1.6truein {\epsffile{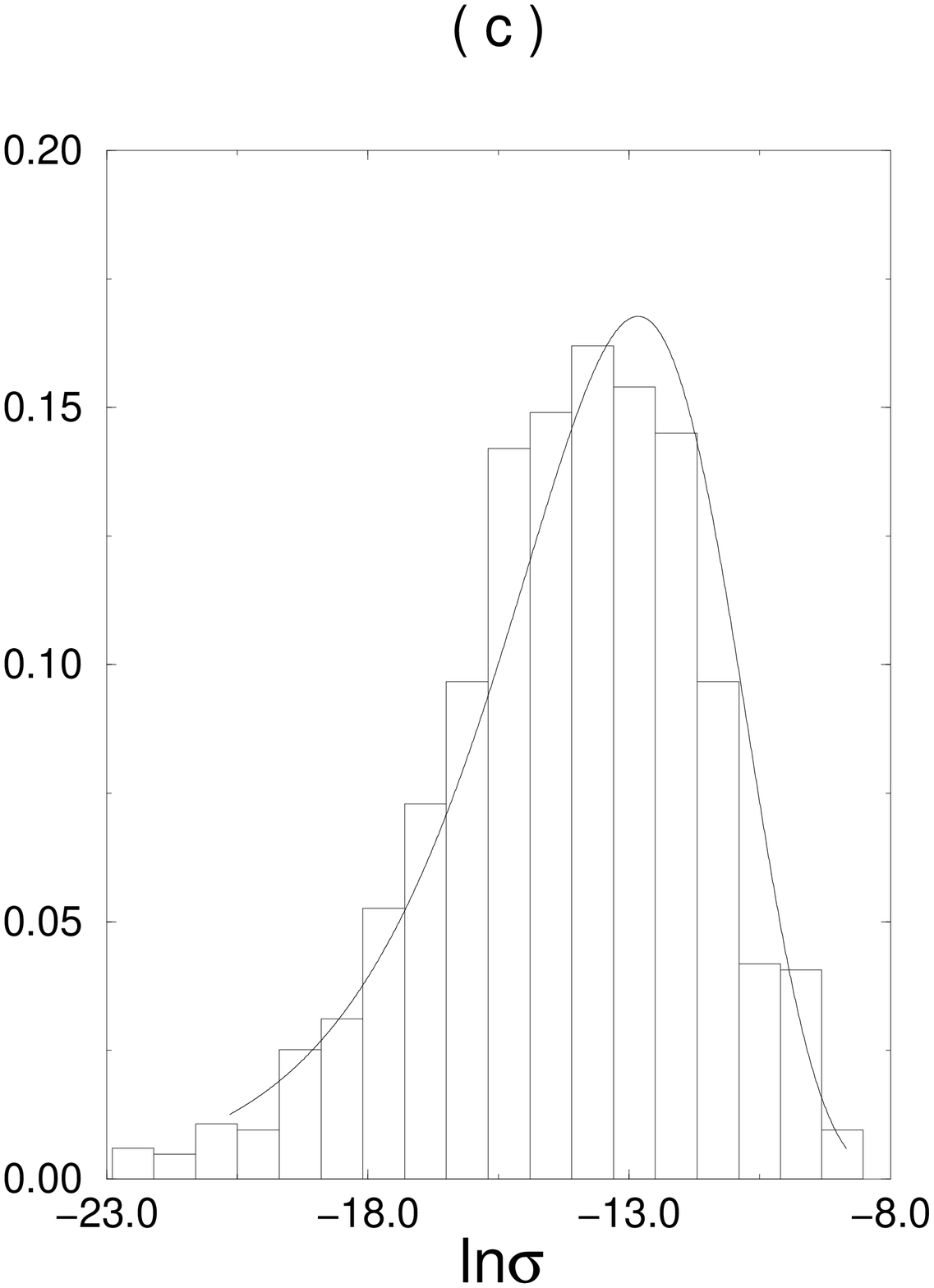}}}
\caption{The ensemble distribution function for the conductance  of 2D system: 
a) thin film; b) square sample; c) long system. Histograms are numerical 
results. The solid 
curve in b) is the Gaussian fitting, in c) is the RR theoretical fitting for the 
1D case.}
\end{figure}

We see that the situations of narrow 2D is close to 
the 1D case as expected, and the normal 2D is  close to Gaussian. But, 
unexpectedly,
the DF for thin film is similar to the 1D case distribution function and not a 
mirror reflection of it, as argued by RR.

In conclusion, the paper has studied the DF of  the conductance in  
mesoscopic systems by numerical simulation. We have found that the distributions
obtained by choosing 
randomly the chemical potentials (for a fixed impurity configuration), which 
corresponds to a typical experimental situation, coincide  with those obtained 
when both impurity configuration and chemical potential is chosen randomly, in 
agreement with the ergodicity hypothesis. The DFs obtained for one-dimensional 
systems were found to be quite close to the predictions of the theory by Raikh 
and Ruzin. 
For $D=2$, the DF both for  narrow system and thin film looks similar (and close 
to the $1D$ case).The distribution function for the conductance of the square 
sample is close to Gaussian.

This research was partially supported by THE ISRAEL SCIENCE FOUNDATION founded by 
The Israel Academy of Sciences and Humanities

\end{multicols}
\end{document}